\documentclass[twocolumn,showkeys,showpacs,preprintnumbers,prd,superscriptaddress,nofootinbib]{revtex4-1}
\usepackage{graphicx}	
\usepackage{amssymb}
\usepackage{dcolumn}
\usepackage{mathtools}
\usepackage{amsmath}
\usepackage{xcolor}
\usepackage{color}
\usepackage{theorem}
\usepackage{subfigure, rotating, bm, array}
\usepackage[pagebackref=false, colorlinks=true]{hyperref}
\hypersetup{linkcolor=blue, citecolor=blue,urlcolor=blue} 

\begin{document}

\title{Influence of primary hair and plasma on intensity distribution of black hole shadows}% Force line breaks with \\

\author{Vitalii Vertogradov}
\affiliation{Physics Department, Herzen State Pedagogical University of Russia, 48 Moika Emb., Saint Petersburg 191186, Russia
SPb branch of SAO RAS, 65 Pulkovskoe Rd, Saint Petersburg 196140, Russia}
\email{vdvertogradov@gmail.com}
\author{Maxim Misyura}
\affiliation{Department of High Energy and Elementary Particles Physics, Saint Petersburg State University, University Emb. 7/9, Saint Petersburg, 199034, Russia}
\email{max.misyura94@gmail.com} 
\author{Parth Bambhaniya}
\affiliation{International Center for Space and Cosmology, Ahmedabad University, Ahmedabad 380009, Gujarat, India}
\email{grcollapse@gmail.com}

\date{\today}

\begin{abstract}

In this paper, we investigate the influence of primary hair ($l$) on the shadows of hairy Schwarzschild and Reissner-Nordström black holes obtained through gravitational decoupling. In the context of hairy Schwarzschild black holes, $l$ either has no effect or consistently enlarges the photon sphere radius. Notably, even when it violates the strong energy condition, it can decrease the radius. For Reissner-Nordström black holes, an additional matter field consistently expands the photon sphere radius, potentially reaching $3M$, akin to the pure Schwarzschild case. Remarkably, we demonstrate that black holes can exist even when overcharged ($Q^2 > M^2$), casting shadows. Specific intensity calculations reveal $l$ consistently reduces it in both scenarios. Furthermore, we investigate the impact of pressureless plasma, finding $l$ exerts a stronger influence on visible size than plasma. These results can help in our understanding of theoretical models of black hole shadows and can be tested by comparison with the images obtained by EHT collaboration.

{\bf keywords:} Black holes, Shadows, EHT, Plasma, No-hair theorem.
\end{abstract}
\maketitle

%\tableofcontents

\section{Introduction}
Black holes possess distinctive features, notably an event horizon, a one-way membrane in spacetime where objects, including light, can only enter but not escape. However, the existence of an event horizon lacks conclusive evidence. The curvature of spacetime near a black hole is so strong that the light can move along a circle. These light rings around black hole form a unstable photon sphere which forms a black hole shadow~\cite{Synge:1966okc}, the images of which have been obtained by the Event Horizon Telescope Collaboration for the central object in M87 galaxy~\cite{EventHorizonTelescope:2019dse,EventHorizonTelescope:2019uob} and our own galaxy~\cite{EventHorizonTelescope:2022xnr}. These images show that a black hole can't be perceived as an exotic theoretical object anymore but it is a real astrophysical one.

Spacetime around a supermassive compact object is generally believed to abide by vacuum solutions of the Einstein field equations, which include Schwarzschild or Kerr solutions. On the other hand, observations indicate toward a dense concentration of matter surrounding the Galactic Center. As consequence, the existence of a vacuum solution in this situation seems less probable. Therefore, it becomes crucial to speculate about a matter distribution around the center while looking at the physical signatures that go along with it. This approach allows it conceivable to test the no-hair theorem and the influence of plasma on black holes in the context of the Galactic Center. A shadow is a very useful tool to test different black hole models and modified theories of gravitation~\cite{Vagnozzi:2022moj}. It can also be used as a standard cosmological ruler ~\cite{Tsupko:2019pzg,Vagnozzi:2020quf}. 

When light from a distant source approaches a black hole, it can be captured by the black hole. The border of escaping to the distant observer is called a photon sphere. Thus, the observer should see the black spot against the light source distribution. A comprehensive review of a black hole shadow can be found in the paper~\cite{Perlick:2021aok}. Various  properties of a black hole shadow have been extensively studied by numerous authors~\cite{Claudel:2000yi,Hod:2012ax,Khoo:2016xqv,Decanini:2010fz,Shoom:2017ril,Cederbaum:2015fra,Johannsen:2013vgc,Abdujabbarov:2015xqa,Younsi:2016azx,Perlick:2018iye,Tsupko:2021yca,Bambhaniya:2021ybs,Bambhaniya:2021ugr,bib:vaidya1,bib:vaidya2,Heydarzade:2023gmd,bib:tsupko_first, bib:tsupko_plasma, bib:tsupko_lensing, bib:understanding, bib:joshi_shadow, bib:Arpan2021,bib:tsukamoto}. Recently, it has been understood that naked singularities can also cast a shadow~\cite{bib:joshi_shadow,Saurabh:2023otl,Saurabh:2022jjv,Solanki:2021mkt}. Therefore, it is not possible to distinguish between a naked singularity and a black hole based on their shadows. 

In general, individuals believe that a photon sphere is what causes the shadows to form. But recently, the authors of \cite{Joshi:2020tlq} introduced a new spherically symmetric null naked singularity solution to the Einstein field equation that casts a shadow albeit not having a photon sphere. They then derive the general conditions under which a shadow could form in the absence of a photon sphere for both timelike and null naked singularities, where both sorts of singularities fulfil all the energy conditions \cite{Joshi:2020tlq}.

A famous fact regarding black holes is the `no-hair' theorem, which
states that a black hole can be described using only three parameters, i.e.
 mass $M$, electric charge $Q$ and spin $a$. However, it has
shown that a black hole may possess a soft hair~\cite{bib:hok_hair}.
Recently, it has been understood that the `no-hair' theorem can be
evaded by using minimal geometrical deformations~\cite{bib:mgd1,bib:mgd2, bib:mgd3} and extended gravitational decoupling methods~\cite{bib:gd1, bib:gd2, bib:gd3}. Ovalle et all~\cite{bib:bh1} have introduced a generalization of a Schwarzschild black hole
surrounded by an anisotropic fluid, which can have primary hair due to extended gravitational decoupling. This new solution has provoked a significant number of papers that have further generalized this solution to hairy Kerr~\cite{bib:hairy_kerr}, Vaidya and generalized Vaidya~\cite{bib:vermax}, regular hairy  black holes~\cite{bib:vermax2,  bib:ovalle_regular} and many others. Gravitational decoupling represents a novel and powerful tool for obtaining new solutions to the Einstein equations.

Gravitational decoupling enables the derivation of new solutions of the Einstein field equations that describe the exterior geometry of a black hole. These compact objects have fascinating physical and geometrical properties, and it would be observationally significant to distinguish them. Therefore, it is crucial to study the observational aspects of these solutions and test them with observational data to confirm or reject a model. This paper explores the properties of the black hole shadow of hairy Schwarzschild and Reissner-Nordstrom black holes obtained by gravitational decoupling. The impact of a primary hair on the radius of the photon sphere and black hole shadow is investigated. Additionally, the specific intensity is calculated, and it is shown that a primary hair always reduces it compared to the usual black hole solution. The influence of pressureless plasma on the size of a shadow is also evaluated.

This paper is organized as follows. In sec. \ref{sec2}, we consider the influence of a primary hair on the black hole shadow in hairy Schwarzschild and Reissner-Nordstrom spacetimes. In sec. \ref{sec3}, we prove that a primary hair can lead to an overcharged black hole with $Q^2>M^2$ ,and this black hole can also cast a shadow. In sec. \ref{sec4}, we evaluate the specific intensity of these solutions and show that a primary hair always decreases it. We investigate the influence of pressureless plasma on a shadow in sec. \ref{sec5}, and sec. \ref{sec6} is the discussions and conclusions. The system of units which we use throughout the paper, is geometrized one in which $G=c=1$. The metric signature is considered as $-+++$.

\section{Shadow of hairy black holes}
\label{sec2}
In this section, we will consider several hairy black hole models suggested in the paper~\cite{bib:bh1}. We express the line element in terms of the lapse function $f=f(r)$
\begin{equation} \label{eq:genmet}
ds^2=-fdt^2+f^{-1}dr^2+r^2d\Omega^2,
\end{equation}
here $d\Omega^2=d\theta^2+\sin^2\theta d\varphi^2$ is the metric on two-sphere.

The spacetime \eqref{eq:genmet} admits two constants of motion, i.e. energy per unit mass $E$
\begin{equation} \label{eq:energy}
E=f\frac{dt}{d\lambda} \,,
\end{equation}
and angular momentum per unit mass $L$
\begin{equation} \label{eq:angular}
L=r^2\sin^2\theta \frac{d\varphi}{d\lambda},
\end{equation}
where $\lambda$ is an affine parameter along the null geodesic. The considered models are spherically symmetric. Therefore, without loss of generality, we can consider the motion only in the equatorial plane $\theta=\frac{\pi}{2}$. When discussing the black hole shadow, the primary objective is to locate the unstable photon orbit, also known as the photon sphere. By using the eqn. \eqref{eq:energy}, \eqref{eq:angular} and the null geodesic condition, $g_{ik}u^iu^k=0$, one can obtains the radial part of geodesic equation
\begin{equation} \label{eq:radial}
\left(\frac{dr}{d\lambda}\right)^2=-V_{eff},
\end{equation}
where
\begin{equation} \label{eq:potential}
V_{eff}=f\frac{L^2}{r^2}-E^2 \,,
\end{equation}
is an effective potential.
The radius $r=r_{ph}$ is the radius of photon sphere, if the following conditions are held:
\begin{equation} \label{eq:condition_potential}
V_{eff}(r_{ph})=0,~~ V'_{eff}(r_{ph})=0 \,,
\end{equation}
where a prime denotes the derivative with respect to the radial coordinate $r$.
Using the second condition from \eqref{eq:condition_potential}, it can determine the radius of the possible photon sphere:
\begin{equation}
V'_{eff}(r_{ph})=0 \rightarrow f'(r_{ph})r_{ph}-2f(r_{ph})=0 \,.
\end{equation}
By using the first condition \eqref{eq:condition_potential} and introducing the impact parameter $b=\frac{L}{E}$, one can obtain the visible size of the shadow,
\begin{equation}
V_{eff}(r_{ph})=0 \rightarrow b=\frac{r_{ph}}{\sqrt{f(r_{ph})}} \,.
\end{equation}
The observer will see the black spot of the radius $b$, where coordinates are,
\begin{eqnarray}
x&=&b\cos \psi,\nonumber \\
y&=&b\sin \psi,\nonumber \\
0&\leq &\psi < 2\pi.
\end{eqnarray}

\subsection{model 1}

The first model which we examine is the hairy Schwarzschild black hole, the lapse function $f$ of which is given by
\begin{equation} \label{eq:hairy1}
f(r)=1-\frac{2m}{r}+\alpha e^{-\frac{r}{M}} \,,
\end{equation}
where the Schwarzschild mass $M$ is related to the new mass parameter $m$ by the relation
\begin{equation}
m=M+\frac{\alpha l}{2} \,.
\end{equation}
Here, $\alpha$ is a coupling constant, $l$ is related to a primary hair, and it has the length dimension. The influence of these parameters on the hawking temperature, geodesic motions, gravitational lensing and energy extraction has been investigated in ~\cite{bib:thermo, bib:kudr_thermo, bib:geod, bib:lens, bib:energy}.
To find out what impact these parameters have on the shadow of a black hole, we use eqn. \eqref{eq:hairy1} in the second condition of eqn. \eqref{eq:condition_potential},
\begin{equation} \label{eq:conditionh1}
V'_{eff}(r_{ph})=0 \rightarrow \frac{6m}{r}-\alpha e^{-\frac{r}{M}}\left(\frac{r}{M}+2\right)-2= 0\,.
\end{equation}
If we assume that the radius of the photon sphere is not influenced by the new matter source then from eqn. \eqref{eq:conditionh1}, we obtain the following value for a primary hair $l$ regardless of the value of $\alpha$.
\begin{equation} \label{eq:key1}
r_{ph}=3M \rightarrow l=\frac{5M}{e^3} \,.
\end{equation}
 However, the visible size of the black hole shadow will change as,
\begin{equation} \label{eq:impacth1}
b=\frac{ 3 \sqrt{3}M }{\sqrt{1-2\alpha e^{-3}}} \,.
\end{equation}
Now, as shown in eqn. \eqref{eq:impacth1}, if there is no additional matter source $\alpha \rightarrow 0$, the visible size of the black hole shadow remains unchanged $b=3\sqrt{3}M$ - as it would be. However, the presence of an extra matter source increases the visible size of the black hole shadow if
\begin{equation}
0 < \alpha <\frac{e^3}{2} \,.
\end{equation}
One should note that the coupling constant is assumed to be less than unity, so the case $\alpha =\frac{e^3}{2} \rightarrow b=\infty$ corresponds to unphysical situation.
If we consider the eqn. \eqref{eq:conditionh1} in general case then the impact parameter is
\begin{equation} \label{eq:impacth11}
    \begin{split}
        b^2=&\frac{r_{ph}^3(r_{ph}+2M)}{r_{ph}^2-\beta r_{ph}+M\beta}, \\\\[-10pt]
\beta =&\alpha l+2M.
    \end{split}
\end{equation}
One should note that both $l$ and $\alpha$ are supposed to be positive. So $\beta>0$ and the black hole shadow exists for $r_{ph}>M$, as can be seen from the denominator of eqn. \eqref{eq:impacth11}. Also, if we assume that $\alpha=0$ then $\beta=2M, r_{ph}=3M$ and $b=3\sqrt{3}M$ - Schwarzschild case.
However, the influence of a primary hair $l$ on the radius of a photon sphere is not clear from eqn. \eqref{eq:conditionh1}. We know that a parameter $\alpha$ is supposed to be small $\alpha \ll 1$, so we can consider the radius of a photon sphere in the form, 
\begin{equation} \label{eq:pert_radius}
r_{ph}=r_{ph}^{(0)}+\alpha r_{ph}^{(1)} \,,
\end{equation}
where $r_{ph}^{(0)}=3M$ is the radius of a photon sphere for the Schwarzschild black hole.
If we apply the first condition of eqn. \eqref{eq:condition_potential} to the effective potential and write it in the form:
\begin{equation} \label{eq:charged_first}
V_{eff}=0 \rightarrow b^2F(r)\left(1+\alpha G(r) \right)=1 \,,
\end{equation}
where,
\begin{eqnarray} \label{eq:def}
F(r)&\equiv &\frac{r-2m}{r^3},\nonumber \\
G(r)&\equiv &\frac{e^{-\frac{r}{M}}-\frac{l}{r}}{1-\frac{2M}{r}}.
\end{eqnarray}
Now, applying the second condition of eqn.\eqref{eq:condition_potential} into \eqref{eq:charged_first}, one can obtains
\begin{equation} \label{eq:charged_second}
\frac{dV_{eff}}{dr}=0 \rightarrow F'(1+\alpha G)+\alpha FG' =0,
\end{equation}
where a prime denotes the derivative with respect to the $r$ coordinate. From eqn. \eqref{eq:charged_second}, one can easily find that $F'(r_{ph}^{(0)})=0$. Substituting eqn. \eqref{eq:pert_radius} into \eqref{eq:charged_second} and expand on small parameter $\alpha$, we can finds
\begin{eqnarray} \label{eq:sol}
&& F''(r_{ph}^{(0)})\alpha r_{ph}^{(1)}+\alpha G'(r_{ph}^{(0)})F(r_{ph}^{(0)})=0 \rightarrow\nonumber \\
&&r_{ph}^{(1)}=-\frac{G'(r_{ph}^{(0)})F(r_{ph}^{(0)})}{F''(r_{ph}^{(0)})}.
\end{eqnarray}
Now, by using eqn. \eqref{eq:def} into \eqref{eq:sol}, we can evaluate the sign of $r_{ph}^{(1)}$ in order to find out whether the primary hair will decreases or increases the radius of a photon sphere. For this purpose, let's find $F(r_{ph}^{(0)})\,, F''(r_{ph}^{(0)})\,, G'(r_{ph}^{(0)})$:
\begin{eqnarray}
F\left(r_{ph}^{(0)}\right)&=&\frac{1}{27M^2},\nonumber \\
F''\left(r_{ph}^{(0)}\right)&=&-\frac{2}{81M^4},\nonumber \\
G'\left(r_{ph}^{(0)}\right)&=&\frac{l}{M^2}-\frac{5}{M}e^{-3}.
\end{eqnarray}
Substituting these results into \eqref{eq:sol}, one obtains for $r_{ph}^{(1)}$
\begin{equation} \label{eq:solution1}
r_{ph}^{(1)}=\frac{3l}{2}-\frac{15M}{2}e^{-3} \,.
\end{equation}
From this result we can note the following points:
\begin{itemize}
\item $l=5Me^{-3} \rightarrow r_{ph}^{(1)}=0$. We arrive at the condition \eqref{eq:key1}, when a primary hair doesn't influence the radius of a photon sphere for the Schwarzschild black hole;
\item $l>5Me^{-3}$. In this case, $r_{ph}>r_{ph}^{(0)}$, i.e. a primary hair $l$ increases the radius of a photon sphere in comparison  with the Schwarzschild black hole;
\item $l<5Me^{-3}$. This case leads to the smaller radius of photon sphere than in Schwarzschild case $r_{ph}<r_{ph}^{(0)}$. However, in this case, the energy-momentum tensor of extra matter fields violates the strong energy condition~\cite{bib:bh1}. 
\end{itemize}
Thus, if one demands the fulfilment of the strong energy condition, then the radius of the photon sphere always increases. Note that all these conditions don't depend upon the coupling constant $\alpha$. The only constraint we impose on it is $\alpha \ll 1$.
\subsection{model 2}

The second model is given by the lapse function $f(r)$
\begin{equation} \label{eq:lapseh2}
f(r)=1-\frac{2M}{r}-\frac{\alpha l}{r}+\frac{Q^2}{r^2}-\frac{\alpha M}{r}e^{-\frac{r}{M}} \,.
\end{equation}
To fulfill the dominant energy condition, one should impose the following restrictions on $Q$ and $l$~\cite{bib:geod}
\begin{equation}
Q^2\geq \frac{\alpha M^2}{e^2},~~ l\geq \frac{M}{e^2} \,.
\end{equation}
If we substitute the lapse function \eqref{eq:lapseh2} into the second condition of eqn. \eqref{eq:condition_potential}, then one obtains
\begin{equation} \label{eq:fconditionh2}
\begin{split}
    V'_{eff}(r_{ph})=0 &\rightarrow \frac{6M+3\alpha l}{r}-\frac{4Q^2}{r^2}-2 \\& +\left(\frac{3\alpha M}{r}+\alpha\right)e^{-\frac{r}{M}} \,.
\end{split}
\end{equation}
In usual Reissner-Nordstrom solution ($\alpha=0$), one has \begin{eqnarray} \label{eq:rn}
2r_{ph}&=&3M+\sqrt{9M^2-8Q^2},\nonumber \\
b&=&\frac{r^2_{ph}}{\sqrt{Mr_{ph}-Q^2}}.
\end{eqnarray}
Note that in the hairy Schwarzschild case, one might have the same $r_{ph}$ as in 
the Schwarzschild case. The comparison of eqn. \eqref{eq:rn} with eqn. \eqref{eq:fconditionh2} shows that there is no photon sphere at $r=r_{ph}$ for the considered hairy model.
If we substitute \eqref{eq:fconditionh2} into the first condition of eqn. \eqref{eq:condition_potential} then one obtains
\begin{eqnarray} \label{eq:sconditionh2}
&&\beta = 3+\frac{r_{ph}}{M}\,, \nonumber \\
&&\kappa=\beta(r_{ph}^2-2 m r_{ph}+Q^2)+ \nonumber \\
&&+6M r_{ph}+3\alpha l r_{ph}-4Q^2-2r_{ph}^2 \,,\nonumber \\
&&b=\frac{\sqrt{\beta}\,r_{ph}^2}{\sqrt{\kappa}}\,.
\end{eqnarray}
Again, like in the previous model, the expression \eqref{eq:fconditionh2} is too cumbersome to find out what an impact a primary hair $l$ has on the radius of a photon sphere. Thus, To evaluate this influence, we utilize the same method as in the previous subsection. We assume that the radius of a photon sphere has the form \eqref{eq:pert_radius} and, using the first condition of eqn. \eqref{eq:condition_potential}, we write down the effective potential in the form of eqn. \eqref{eq:charged_first}. The functions $F(r)$ and $G(r)$ for the spacetime with lapse function \eqref{eq:lapseh2} are given by,
\begin{eqnarray} \label{eq:def2}
r_{ph}^{(0)}&=&\frac{1}{2}\left(3M+\sqrt{9M^2-8Q^2}\right),\nonumber \\
F(r)&=&\frac{r^2-2Mr+Q^2}{r^4},\nonumber \\
G(r)&=&-\frac{r\left(l+Me^{-\frac{r}{M}}\right)}{r^2-2Mr+Q^2}.
\end{eqnarray}
Now applying the same method as described above, we evaluate functions $F$, $F''$ and $G$ at the radius $r_{ph}^{(0)}$, and then we have
\begin{eqnarray}
F\left(r_{ph}^{(0)}\right)&=&\frac{Mr_{ph}^{(0)}-Q^2}{\left(r_{ph}^{(0)}\right)^4},\nonumber \\
F''\left(r_{ph}^{(0)}\right)&=&\frac{8Q^2-6Mr_{ph}^{(0)}}{\left(r_{ph}^{(0)}\right)^6},\nonumber \\
G'\left(r_{ph}^{(0)}\right)&=&\frac{3l+3Me^{-\frac{r_{ph}^{(0)}}{M}}+r_{ph}^{(0)}e^{-\frac{r_{ph}^{(0)}}{M}}}{Mr_{ph}^{(0)}-Q^2},
\end{eqnarray}
substituting these results into the definition $r_{ph}^{(1)}$ \eqref{eq:sol}, we can obtain
\begin{equation} \label{eq:radius1h2}
r_{ph}^{(1)}=-\left(r_{ph}^{(0)}\right)^2 \frac{3l+3Me^{-\frac{r_{ph}^{(0)}}{M}}+r_{ph}^{(0)}e^{-\frac{r_{ph}^{(0)}}{M}}}{8Q^2-6Mr_{ph}^{(0)}} \,.
\end{equation}
As one can see that the sign of $r_{ph}^{(1)}$ depends on the sign of the expression
\begin{equation} \label{eq:imh2}
8Q^2-6Mr_{ph}^{(0)}=8Q^2-9M^2-3M\sqrt{9M^2-8Q^2} \,.
\end{equation}
Introducing a dimensionless parameter $\xi$ by
\begin{equation}
\xi^2=\frac{Q^2}{M^2},~~ 0\leq \xi \leq 1 \,,
\end{equation}
one can write eqn. \eqref{eq:imh2} as,
\begin{equation}
-9M^2\left(1-\frac{8}{9}\xi^2+\sqrt{1-\frac{8}{9}\xi^2}\right)<0 \,.
\end{equation}
Thus, we can come to the conclusion that in the hairy Reissner-Nordstrom case, the primary hair $l$ always increases the radius of a photon sphere in comparison  with the RN black hole.
The radius of a photon sphere can be equal to $3M$ like in Schwarzschild spacetime. In this case, from eqn. \eqref{eq:fconditionh2}, we obtain
\begin{equation}
r=3M \rightarrow l=\frac{4Q^2}{9\alpha M}-2e^{-3} \,.
\end{equation}

\section{Shadow of overcharged black hole}
\label{sec3}
In pure Reissner-Nordstrom spacetime, if a black hole is overcharged, i.e. $Q^2>M^2$, then this spacetime contains a naked singularity. It doesn't cast a shadow. However, the primary hair $l$ affects the spacetime and overcharged black holes with $Q^2>M^2$ can exist. We will prove that overcharged black holes can also cast a shadow. We will show it implicitly since the lapse function \eqref{eq:lapseh2} and the second condition of eqn. \eqref{eq:fconditionh2} are too cumbersome to obtain an analytical solution. First, we will show that there can be a black hole with $Q^2>M^2$, and then we will prove that one of the radii of the photon spheres is outside the event horizon and the other is inside. Here, we consider the charge $Q$ as
\begin{equation} \label{eq:charge_pert}
Q^2 = M^2+ \delta,~~ \delta \ll M^2 ,
\end{equation}
then the lapse function \eqref{eq:lapseh2} can be written as,
\begin{equation} \label{eq:lapsehext}
f(r)=\left(1-\frac{M}{r}\right)^2+\frac{\delta}{r^2}-\frac{\alpha l}{r}-\frac{\alpha M}{r}e^{-\frac{r}{M}} \,.
\end{equation}
From eqn. \eqref{eq:lapsehext}, it is obvious that we can choose parameters $\delta\,, \alpha \,, l$ such that $f(M)<0$ and $f(2M)>0$, which indicates the existence of the event horizon in the region $M <r<2M$. 
Now, by substituting the charge \eqref{eq:charge_pert} into the second condition of eqn. \eqref{eq:fconditionh2}, one obtains:
\begin{equation} \label{eq:naked_shadow}
    \begin{split}
      G(r) &\equiv \frac{6M+3\alpha L}{r}-\frac{4Q^2+4\delta}{r^2}- \\ &2+\left(\frac{3 \alpha M}{r}+\alpha\right)e^{-\frac{r}{M}} =0 \,. 
    \end{split}
\end{equation}
Neglecting $\frac{\delta}{M^2}$ in comparison with $(\frac{3\alpha l}{2M}+\left(\frac{3\alpha l}{2M}+\alpha\right) e^{-2})$\footnote{To obtain this expression, we put $r=2M$ into the expression of eqn. \eqref{eq:naked_shadow}.}, it is easy to show that $G(M)$ and $G(2M)$ are both positive indicating that the first photon sphere radius is the inside the event horizon. 
However, if one satisfies the condition
\begin{equation}
\frac{4}{9}\geq \frac{\alpha l}{M}+\left(\frac{\alpha l}{M}+\alpha \right) e^{-3} \,,
\end{equation}
then the radius of the second photon sphere is located outside the event horizon in the region $2M<r<3M$ and can form a black hole shadow with an impact parameter of the form
\begin{equation}
b=\frac{r_{ph}}{\sqrt{f\left(r_{ph}\right)}}.
\end{equation}
Thus, we implicitly showed that overcharged hairy Reissner-Nordstrom black hole can cast a shadow.

\section{Specific intensity of a shadow}
\label{sec4}
\begin{figure*}[!ht]
 \includegraphics[scale=0.7]{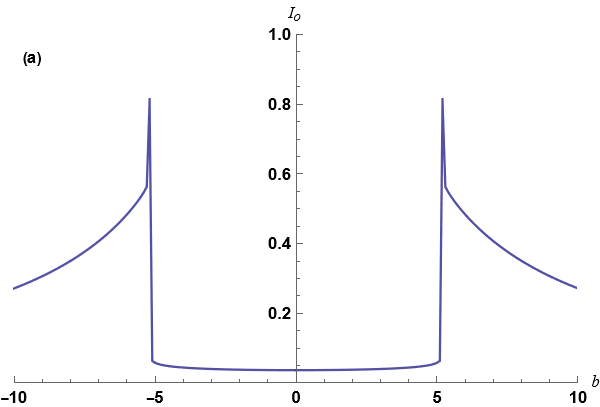}
 \includegraphics[scale=0.7]{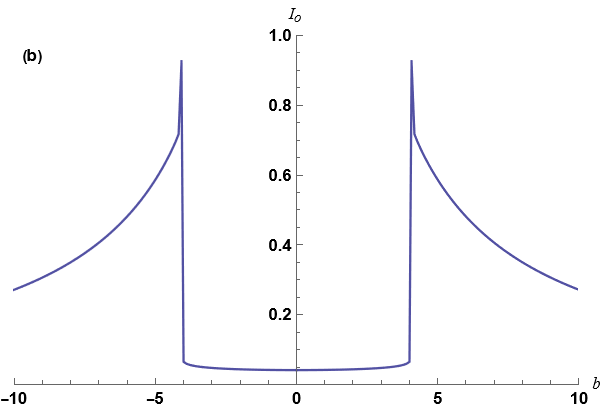}
 \includegraphics[scale=0.7]{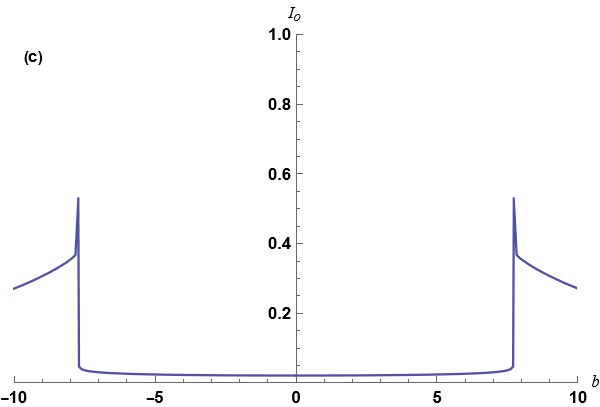}
\includegraphics[scale=0.7]{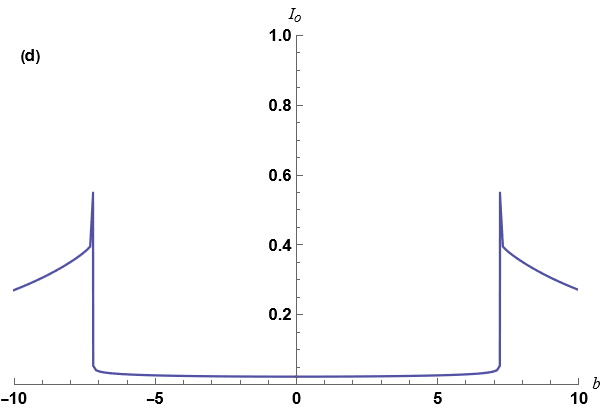}
\centering
    \caption[fig1]{These plots show observed
    intensity distributions as a function of the impact parameter $b$ for $M=1$. Plot (a) shows the function $I_{0}$ versus the impact parameter $b$ for the Schwarzschild black hole.  Plot (b) shows the function $I_{0}$ versus the impact parameter $b$ for the Reissner-Nordstrom black hole with $Q=0.98$. Plot (c) shows the function $I_{0}$ versus the impact parameter $b$ for model $1$ if $\alpha=0.5$ and $l=2$.  Plot (d) shows the function $I_{0}$ versus the impact parameter $b$ for model $2$ if $Q=0.98$, $\alpha=0.5$ and $l=2$.  }
    \label{fig:1}
\end{figure*}

In the previous section, we have considered an idealized scenario where the light source is located far behind a black hole, and there are no light sources between the observer and the compact object. A real astrophysical black hole is typically surrounded by an accretion disc. We will compute the influence of plasma on the radius of a photon sphere in the next section. In this section, we consider an optically thin, radiating, accreting flow that surrounds the hairy black hole and intensity distributions as a function of the impact parameter. The radiation can be emitted in the vicinity of a black hole, including the region inside the photon sphere. This radiation will affect the specific intensity $I_{\nu_0}$ at the observed photon frequency $\nu_0$ in the observer's sky at the point $(x,y)$. It  has the following form~\cite{bib:jaros, bib:bambi}
\begin{equation} \label{eq:intensity}
I_{\nu_0}\left(x,y\right)=\int_{\gamma}G'^3J\left(\nu_e\right)dl_{prop},
\end{equation}
here, $\nu_e$ is emitted frequency, $G'=\frac{\nu_0}{\nu_e}$ is the redshift factor, $J\left(\nu_e\right)$ is the emitter’s rest frame emissivity per unit volume, $dl_{prop}=-K_iu^i_{(e)}d\lambda$ is the infinitesimal proper length in the rest frame of the emitter, $K^i\equiv \frac{dx^i}{d\lambda}$ is the photon four-momentum, ($K^iK_i=0$), $u^i_{(e)}$ is the four-velocity of an emitter and $\lambda$ is an affine parameter. The integration is evaluated along the observed photon path $\gamma$. The redshift $G'$ in terms of the four-velocities $K^i, u^i_{(0)}, u^i_{(e)}$ is given by~\cite{bib:hartle}
\begin{equation} \label{eq:redshift}
G'=\frac{K_iu^i_{(0)}}{K_ju^j_{(e)}} \,,
\end{equation}
where $u^i=\left \{1\,, 0\,, 0\,, 0\right\} $ is the four-velocity of an observer. We consider accreting gas which is in radial free fall with four-velocity $u^i_{(e)}$ of the form~\cite{bib:bambi}
\begin{eqnarray} \label{eq:gas}
u^0_{(e)}&=&\frac{1}{f},\nonumber \\
u^1_{(e)}&=&-\sqrt{1-f},\nonumber \\
u^2_{(e)}&=&u^3_{(e)}=0.
\end{eqnarray}
The four-vector $K^i\equiv \frac{dx^i}{d\lambda}$ is given by:
\begin{eqnarray} \label{eq:photon}
K^0&=&\frac{E}{f},\nonumber \\
K^1&=&\pm\sqrt{E^2-f\frac{L^2}{r^2}},\nonumber \\
K^2&=&0,\nonumber \\
K^3&=&\frac{L^2}{r^2}.
\end{eqnarray}
Here the sign $K^1$ depends on the photon motion, i.e. $+$ if it goes away from a black hole and $-$ if approaches it. Substituting \eqref{eq:photon} and \eqref{eq:gas} into \eqref{eq:redshift}, we find the redshift $G'$ in the form
\begin{equation}
G'=\frac{K_0f}{K_0-fK_1\sqrt{1-f}} \,.
\end{equation}
For the specific emissivity $J$, we assume the following simple model~\cite{bib:bambi} in which the emission is monochromatic with emitter's rest-frame frequency $\nu^*$ and the emission has a $\frac{1}{r^2}$ radial profile:
\begin{equation} \label{eq:emis}
J(\nu_e)\propto \frac{\delta(\nu_{e}-\nu^*)}{r^2} \,.
\end{equation}
The proper length $dl_{prop}$ is given by
\begin{equation} \label{eq:proper}
dl_{prop}=-\frac{-K_0}{G'K^1}dr \,.
\end{equation}
Now substituting \eqref{eq:redshift}, \eqref{eq:emis} and \eqref{eq:proper} into \eqref{eq:intensity} and integrating over all observed frequencies, we find the observed intensity
\begin{equation} \label{eq:intensity_final}
 I_{o}(x,y)\sim -\int_{\gamma}\frac{G'^3K_0}{r^2K^1}dr \,.
\end{equation} 
As we have shown, a primary hair increases the radius of a photon sphere. As a result, the specific intensity decreases, as shown in figures \ref{fig:1} (c) and (d).

\section{The influence of a plasma on a radius of a photon sphere}
\label{sec5}
In an idealized situation, we assume that there is no matter between the observer and the black hole. However, as we showed above, the accretion disc has an impact on the specific intensity. Here we consider the influence of plasma for a radius of a photon sphere of a hairy black hole. Our analysis goes close to the method described in~\cite{bib:tsupko_plasma}. We consider low-density, non-magnetized pressureless plasma. In the absence of the pressure, the particles move along timelike geodesics for which $g_{ik}u^iu^k=-1$. In order to estimate the influence of plasma, one needs to calculate the plasma frequency $\omega_p$ which, for our model, depends on the plasma density $\rho$. Let's consider the continuity equation
\begin{equation} \label{eq:cont}
\left(\sqrt{-g}\rho u^i_{(p)}\right)_{,i}=0 \,.
\end{equation}
Here $g=\det g_{ik}$ and $u^i_{(p)}$ is the four-velocity of the particles. We simplify our model by assuming that electron and proton four-velocities are coincides. The plasma is a neutral hydrogen one. We neglect the electron mass $m_e$ in comparison with proton mass $m_p$. So, the plasma density $\rho$ can be written as
\begin{equation}
\rho \sim m_pN \,.
\end{equation}
Where $N$ is the number density of protons. Again, we assume the equatorial geodesic motion i.e. $\theta=\frac{\pi}{2}$. As a result the metric tensor components depend only upon the radial coordinate $r$ and the eqn. \eqref{eq:cont} becomes
\begin{equation}
\left(r^2\rho u^1\right)_{,r}=0 \,.
\end{equation}
An integration gives,
\begin{equation} \label{eq:integration}
4\pi \rho r^2u^1=-\dot{m}=const. \,,
\end{equation}
where $\dot{m}$ is a stationary mass flux. 
Here, we assume that particles starts its radial motion from rest at infinity. It means $E=1$, then $u^1$ reads
\begin{equation} \label{eq:radialfall}
u^1=-\sqrt{1-f} \,.
\end{equation}
The density $\rho$ can be expressed through the ratio of plasma $\omega_p$ and photon $\omega_0$ frequencies
\begin{equation} \label{eq:formua}
\frac{\omega_p^2}{\omega_0}=\frac{4\pi e^2N}{m_e\omega_0^2}= \frac{ 4 \pi e^2 \rho}{m_p m_e\omega_0^2}=\beta_0\frac{1}{r^2\sqrt{1-f}} \,,
\end{equation}
where,
\begin{equation}
\beta_0=\frac{e^2\dot{m}}{m_em_p\omega_0^2}.
\end{equation}
As we have mentioned above, we consider a low density plasma i.e. $\omega_p\ll\omega_0$. In this case, the angular size $(\theta_{sh})$ of the shadow observed at the radius $r_o$ is given by (See (56) in~\cite{bib:tsupko_plasma})
\begin{equation} \label{eq:angular_plasma}
\begin{split}
     \sin^2 \theta_{sh} & \approx \frac{r_{ph}^2}{\left(r_{o}^{(0)}\right)^2}\frac{f\left( r_o\right)}{f\left(r_{ph}^{(0)}\right)} \biggl( 1-\beta_0f\left(r_{ph}^{(0)}\right) \times \\ & \sqrt{1-f\left(r_{ph}^{(0)}\right)}+\beta_0f(r_o)\sqrt{1-f(r_o)}\biggr) \,.
\end{split}
\end{equation}
Where $r_{ph}^{(0)}$ is the radius of a photon sphere without plasma. 
For estimating the plasma correction numerically, we re-express all quantities in Gaussian CGS units, i.e., we restore factors of $c$ and $G$ . Then $\beta_0$ reads
\begin{equation} \label{eq:beta}
\beta_0=\frac{e^2\dot{m} c^3}{\sqrt{2GM} (GM)^{3/2} m_em_p\omega_0^2}\,.
\end{equation}
The accretion rate $\dot{m}$ can be estimated through observed luminosity as,
\begin{equation}
\lambda \sim \eta \dot{m}c^2 \,,
\end{equation}
where $\eta$ is dimensionless quantity related to accretion efficiency. An observed photon frequency at infinity is related to the wavelength $\lambda_0$ by relation,
\begin{equation}
\omega_0=\frac{2\pi c}{\lambda_0} \,.
\end{equation}
We consider two models $\eta=10^{-4}$ and $\eta=0.1$. and two massive black holes Sgr $A^*$ and M87*.
\begin{equation}
\begin{split}
\text{Sgr $A^*$} \quad, \quad M= 4.3\times 10^6 \, M_{\odot} \,, \\
r_o=8.3 \,Kpc \,, \\
\text{M87*} \quad, \quad M=3\times 10^9 M_{\odot} \,, \\
r_o=18\, Mpc \,, \\
\lambda=10^6 \lambda_{\odot} \,, \lambda_0=1 \,\text{mm}\,, \lambda_0=10\, \text{cm}\,,
\end{split}
\end{equation}
without plasma, the angular radius of the shadow of black hole Sgr $A^*$ has the value $\theta_{sh} \simeq 26.572$ $\mu\text{as}$. In general, as one can see from tables \ref{tab:table1} and \ref{tab:table2}, the influence of a primary hair on the visible size of a shadow is much stronger than the influence of plasma except for several particular cases of large $\lambda_0$ and small $\eta$. For pure M87*, the angular radius of the shadow has the following value $\theta_{sh} \simeq  8.548$ $\mu\text{as}$, and the effect of plasma gives a change in the size of the shadow of the order $10^{-6}  \mu \text{as}$  for $\eta= 10^{-4}$ and $\lambda_0 = 10$, whereas primary hair for $\alpha=0.5$ and $l=1.5$ gives the value $\theta_{sh} \simeq 11.602 \mu\text{as}$.

\begin{table}[h!]
\caption{\label{tab:table1} This table shows the value of the angular radius of the shadow for Sgr $A^*$ black hole for different quantities  of parameters $\alpha$ and $l$. The Upper part corresponds to the Schwarzschild case, the middle part to the Reissner-Nordstrom black hole with $Q=0.1$ and the bottom of the table for $Q=0.9$. The units of measure are $\mu\text{as}$ }
\begin{ruledtabular}
\begin{tabular}{cccc}
  Schwarzschild   & $l=0.5 $ & $l=1$ & $l=1.5$ \\ \hline
 $\alpha= 0.01$  & $26.618$ & $26.685$ & $26.751$ \\
 $\alpha= 0.1$  & $27.044$ & $27.718$ & $28.391$ \\
 $\alpha= 0.5$  & $29.052$ & $32.583$ & $36.066$  \\ 
\hline   \hline \\[-3pt]
  RN $Q=0.1$ & $l=0.5 $ & $l=1$ & $l=1.5$ \\ \hline
 $\alpha= 0.01$  & $26.601$ & $26.667$ & $26.734$ \\
 $\alpha= 0.1$  & $27.254$ & $27.915$ & $28.576$ \\
 $\alpha= 0.5$  & $30.074$ & $33.332$ & $36.61$  \\
 \hline   \hline \\[-3pt]
   RN $Q=0.9$  & $l=0.5 $ & $l=1$ & $l=1.5$ \\ \hline
 $\alpha= 0.01$  & $22.190$ & $22.274$ & $22.359$ \\
 $\alpha= 0.1$  & $23.078$ & $23.887$ & $24.684$ \\
 $\alpha= 0.5$  & $26.594$ & $30.248$ & $33.839$  \\
\end{tabular}
\end{ruledtabular}
\end{table}

\vspace{60mm}
\begin{table}[h!]
\caption{\label{tab:table2} This table shows the influence of plasma on the angular radius of the shadow for Sgr $A^*$ black hole for different parameters  $\eta$ and $\lambda_0$. The Upper part corresponds to the pure Schwarzschild case, the middle part to the pure Reissner-Nordstrom black hole with $Q=0.1$, and the bottom of table for $Q=0.9$. The units of measure are $\mu\text{as}$ }
\begin{ruledtabular}
\begin{tabular}{ccc}
 Schwarzschild   & $\eta=10^{-4} $ & $\eta=0.1$   \\ \hline
 $\lambda_0= 10$  & $22.4$ & $26.568$   \\
 $\lambda_0= 0.1$  & $26.571$ & $26.572$   \\ 
 \hline   \hline \\[-3pt] 
  RN $Q=0.1$  & $\eta=10^{-4} $ & $\eta=0.1$   \\ \hline
 $\lambda_0= 10$  & $22.367$ & $26.524$   \\
 $\lambda_0= 0.1$  & $26.527$ & $26.527$   \\ 
 \hline   \hline \\[-3pt] 
   RN $Q=0.9$ & $\eta=10^{-4} $ & $\eta=0.1$   \\ \hline
 $\lambda_0= 10$  & $19.077$ & $22.085$   \\
 $\lambda_0= 0.1$  & $22.087$ & $22.087$   \\
\end{tabular}
\end{ruledtabular}
\end{table} 

 \vspace{20mm}
\section{Discussions and Conclusions}
\label{sec6}
In this paper, we have considered the influence of a primary hair $l$ on the shadow of hairy Schwarzschild and Reissner-Nordstrom black holes obtained by gravitational decoupling. We have found out that in the hairy Schwarzschild case, a primary hair $l$ either doesn't have any impact or always increases the radius of the photon sphere. It can decrease the radius of a photon sphere, but in this case, the energy-momentum tensor $\Theta_{ik}$ violates the strong energy condition. For the Reissner-Nordstrom case, we have shown that an additional matter field $\Theta_{ik}$ always increases the radius of a photon sphere, and this radius can even be equal to $3M$ like in the pure Schwarzschild case.
We also proved that a black hole can exist even in an overcharged case, i.e. when $Q^2>M^2$ and it also casts a shadow. We have calculated the specific intensity and showed that a primary hair $l$ always decreases it in both Schwarzschild and Reissner-Nordstrom cases. The influence of a pressureless plasma on the black hole has been also considered. We have found out that, in general, a primary hair $l$ has a stronger influence on the visible size in comparison with pressureless plasma.
However, the real astrophysical black holes are supposed to be rotating ones. In ~\cite{bib:geod}, it has been shown that under some conditions on $\alpha$ and $l$, the hairy Schwarzshild solution can be a mimicker of a Kerr black hole. In the case of a shadow, it can be viewed as a Kerr black hole mimicker only if we observe a shadow on a certain angle. Because, in general the shadow in Kerr spacetime is flattened, but for the black holes considered in this paper, a shadow is always circular in shape. The results obtained in this paper, can be further used for the construction of a black hole shadow for rotating hairy Kerr and Kerr-Newman spacetimes obtained by gravitational decoupling~\cite{bib:hairy_kerr}. After this, all these models can be tested by comparison with the images obtained by EHT collaboration.

{\bf Acknowledgments:} V. Vertogradov thanks the Basis Foundation (grant number 23-1-3-33-1) for the financial support. M. Misyura gratefully acknowledges partial support from the Theoretical Physics and Mathematics Advancement Foundation BASIS, grand no.20-1-5-109-1.

{\bf Data Availability Statement:} No Data associated in the manuscript.

\end{document}